\documentstyle[twoside,fleqn,espcrc2]{article}

\def\beq{\begin{eqnarray}}
\def\eeq{\end{eqnarray}}

\newcommand{\AmS}{{\protect\the\textfont2
  A\kern-.1667em\lower.5ex\hbox{M}\kern-.125emS}}

\title{Mixing and Decay Constants of Pseudoscalar Mesons:\\
Octet-Singlet vs.\ Quark Flavor Basis}

\author{Thorsten Feldmann\thanks{Supported by {\it Deutsche
      Forschungsgemeinschaft}.}
\address{Department of Theoretical Physics, University of Wuppertal,
    D-42097 Wuppertal, Germany\\[0.5em]
{\small
Contribution to the conference {\it QCD 98}\/, Montpellier 2-8 July
1998.\\
\tt hep-ph/9807367, Wuppertal preprint WUB 98-25}}}

\begin{document}

\begin{abstract}
Although $\eta-\eta'$ mixing is qualitatively well understood as
a consequence of the $U(1)_A$ anomaly in QCD together with a
broken $SU(3)_F$ flavor symmetry, until recently the values of decay and
mixing parameters of the $\eta$ and $\eta'$ were only approximately
known, e.g.\ values for the octet-singlet mixing angle between
$-20^\circ$ and $-10^\circ$ could be found in the literature.
New experimental data, especially for the reactions $\gamma\gamma^*
\to \eta,\eta'$ and $B\to \eta'K$, together with new theoretical
results from higher order corrections in chiral perturbation theory
stimulated a phenomenological re-analysis of this subject, which
led to a coherent qualitative and quantitative picture of $\eta-\eta'$
mixing and even of $\eta-\eta'-\eta_c$ mixing.
\end{abstract}

\maketitle

\setcounter{footnote}{0}

\section{$\eta-\eta'$ Mixing Schemes}

A crucial observation of our analysis \cite{FeKrSt98} is the
fact that for a proper treatment of the mixing
one clearly has to distinguish between 
matrix elements of $\eta,\eta'$ states with local currents
(e.g.\ weak decay constants) and overall state
mixing.
While in the former the $SU(3)_F$ symmetry breaking effects,
($2 m_s/(m_u+m_d)
\simeq 26$) turn out to be essential, 
in the latter the gluon anomaly plays the
important role \cite{general}. 
Correspondingly, one may think of two possible
choices of appropriate basis states as a starting
point for the description of $\eta-\eta'$ mixing,
namely the quark flavor basis (which becomes exact in the limit
$m_s \to \infty$) and the octet-singlet basis (which becomes exact
for $m_u=m_d=m_s$), respectively.

In order to define these bases properly, it is useful to consider
a Fock state decomposition of the mesonic states in the
parton picture. One then defines
the quark flavor basis through
\beq
 |\eta\phantom{{}'}\rangle &=& \cos\phi \, |\eta_q\rangle - \sin\phi \,
 |\eta_s\rangle , \cr
 |\eta'\rangle &=& \sin\phi \, |\eta_q\rangle + \cos\phi \,
 |\eta_s\rangle 
\label{eq1}
\eeq
with $\phi$ being the mixing-angle and
\beq
 |\eta_q \rangle &:=&  \Psi_q^{} \,
         |u\bar u + d\bar d\rangle/\sqrt2 + \Psi_q^g \, |gg\rangle + \ldots
         \cr
 |\eta_s \rangle &:=& \Psi_s^{} \,
         |s\bar s \rangle + \Psi_s^g \, |gg\rangle 
  + \ldots
\label{eq2}
\eeq
Here $\Psi_i^n$ denote (light-cone) wave functions of the corresponding
parton states. The effect of higher Fock
states\footnote{Of course, to construct the wave
functions of all Fock states explicitly, one has to solve
the QCD bound state problem.}
($|gg\rangle +
\ldots$) is twofold: First, they are necessary for 
the correct normalization, $\langle\eta_i|\eta_j\rangle
= \delta_{ij}$. Secondly, they reflect the mixing (e.g.\ through the
twist-4 $|gg\rangle$ component 
which is present due to the anomaly).

Analogously, in the octet-singlet basis, one obtains
\beq
 |\eta\phantom{{}'}\rangle &=& \cos\theta \, |\eta_8\rangle - \sin\theta \,
 |\eta_1\rangle , \cr
 |\eta'\rangle &=& \sin\theta \, |\eta_8\rangle + \cos\theta \,
 |\eta_1\rangle 
\label{eq3}
\eeq
with the usual pseudoscalar
octet-singlet mixing angle $\theta=\phi-\arctan\sqrt2$.
However, the flavor decomposition in the Fock state expansion looks
now more complicated due to the broken $SU(3)_F$ symmetry
\beq
 |\eta_8 \rangle &:=&
 \left( \Psi_q^{} \,
         |u\bar u + d\bar d\rangle - 2 \Psi_s^{} \, |s\bar s\rangle
 \right)/\sqrt6 + \cr
&& (\Psi_q^g - \sqrt2 \, \Psi_s^g) \, |gg\rangle /\sqrt3 + \ldots \cr
|\eta_1 \rangle &:=& 
 \left( \Psi_q^{} \,
         |u\bar u + d\bar d\rangle + \phantom{2} \Psi_s^{} \, |s\bar s\rangle
 \right)/\sqrt3 +  \cr
&& (\sqrt2 \Psi_q^g + \Psi_s^g)\, |gg\rangle/\sqrt3 + \ldots \cr &&
\label{eq4}
\eeq
Only in the flavor symmetry limit one would have
trivial relations between the
wave functions,
$\Psi_q^{}=\Psi_s^{}=\Psi_8^{}=\Psi_1^{}$,
$\Psi_q^g = \sqrt2 \Psi_s^g = \sqrt2 \Psi_1^g/\sqrt3$,
$\Psi_8^g = 0$, etc. Only in this case one would
recover the usually anticipated form of octet and singlet
states  
$|\eta_8\rangle \to \Psi_8^{} |u\bar u + d \bar d - 2 s\bar s\rangle
/\sqrt6 + \ldots$ and 
 $|\eta_1\rangle \to \Psi_1^{} |u\bar u + d \bar d + s\bar
 s\rangle/\sqrt3 
+ \Psi_1^g |gg\rangle
+ \ldots$

Note that in higher Fock states with increasing number of partons
the effect of $SU(3)_F$ symmetry breaking is washed out (e.g.\ the
ratio of {\em constituent} quark masses is only $2 \tilde m_s/(\tilde m_u+
\tilde m_d) \approx 5/3$), and thus the octet-singlet basis is
still useful for low-energy expansions of QCD like e.g.\ chiral
perturbation theory (ChPT). However, weak decay constants 
only probe the short-distance properties of
the valence Fock states and are thus rather sensitive to $SU(3)_F$
breaking effects. To see this in more detail, let us define the
decay constants\footnote{We stress that  occasionally used
decay constants
``$f_\eta,f_{\eta'}$'' are
ill-defined quantities.
} as ($f_\pi = 131$~MeV)
\beq
&& \langle 0 | J_{\mu5}^i | P\rangle  \equiv 
\imath \, f_P^i \, p_\mu
\label{eq5}
\eeq
with
$ P=\eta,\eta';\  i=q,s\ (i = 8,1)$, 
and the relevant flavor combinations
of axial-vector currents denoted as $J_{\mu5}^i$.
Using Eqs.~(\ref{eq1}--\ref{eq5}) one obtains
\beq
\left( \begin{array}{cc} f_\eta^q & f_\eta^s \cr f_{\eta'}^q & f_{\eta'}^s
\end{array} \right) &=&  \left( \begin{array}{cc} 
f_q \, \cos\phi & -f_s \, \sin\phi \cr
f_q \, \sin\phi & \phantom{-} f_s \, \cos\phi \end{array} \right) \,
\nonumber \\[0.2em] &=& U(\phi) \, {\rm diag}[f_q,f_s]
\eeq
with $f_q(f_s)$ related to the wave function $\Psi_q^{}(\Psi_s^{})$
at the origin\footnote{\label{foot4}
The decay constants are calculated from
the Fock state decomposition as follows (for concreteness we chose
the $|\eta_s\rangle$ state as an example)
\beq
&&  \langle 0 | J_{\mu5}^s | \eta_s(p)\rangle 
= \sum_{n} 
\int \Psi_s^n \, \langle 0 | J_{\mu5}^s |n(p) \rangle \cr
&=& \int \frac{dx \, d^2k_\perp}{16 \pi^3} \, \Psi_s^{}(x,k_\perp) \,
 \langle 0 | J_{\mu5}^s |s\bar s \rangle \cr
&=& \sum_{\alpha \bar \alpha}
\frac{\delta_{\alpha\bar\alpha}}{\sqrt{N_c}}
\int \frac{dx \, d^2k_\perp}{16 \pi^3} \, \Psi_s(x,k_\perp) \,
\frac{-\imath}{\sqrt2}  {\rm tr}[\gamma_\mu\gamma_5  p
\hskip-0.5em
\slash  \gamma_5]
\cr
&=& \imath \, 2 \sqrt{2 N_c} 
 \int \frac{dx \, d^2k_\perp}{16 \pi^3} \, \Psi_s(x,k_\perp) \, p_\mu
 = \imath  f_s  p_\mu
\label{origin}
\eeq
Here $x$ denotes the usual (light-cone~+) momentum fraction of
the quark and $k_\perp$ its transverse momentum. Note that only the
leading quark-antiquark Fock state contributes to the decay constant,
i.e.\ Eq.~(\ref{origin}) is exact.}, and with $U$ 
being a usual rotation matrix identical to the one of the
state mixing (\ref{eq1}). 

In the octet-singlet basis one obtains on the other hand
\beq
\left( \begin{array}{cc}f_\eta^8 & f_\eta^1 \cr f_{\eta'}^8 & f_{\eta'}^1
\end{array}\right) &=&  \left( \begin{array}{cc} 
f_8 \,\cos\theta_8 & -f_1 \, \sin\theta_1 \cr
f_8 \,\sin\theta_8 & \phantom{-} f_1 \, \cos\theta_1 \end{array} \right) \,
\nonumber \\[0.2em] &\neq&  U(\theta) \, {\rm diag}[f_8,f_1]
\label{tas}
\eeq
where we introduced the parametrization of \cite{Leutwyler97}
\beq
\theta_8 &=& \phi - \arctan\frac{\sqrt2 f_s}{f_q}, \quad f_8^2 =
\frac{f_q^2+2f_s^2}{3} , \cr
\theta_1 &=& \phi - \arctan\frac{\sqrt2 f_q}{f_s}, \quad f_1^2 =
\frac{2f_q^2+f_s^2}{3} 
\label{eqrel}
\eeq
Note that the decay constants do not simply follow the
state mixing in the octet-singlet basis;
-- only in the $SU(3)_F$ symmetry limit one has
$\theta_8 \to \theta \leftarrow \theta_1$.
Especially the matrix elements of octet/singlet currents
with the opposite states do {\em not} vanish,
 $\langle 0 | J_{\mu5}^1|\eta_8\rangle
= \imath p_\mu \, \sin(\theta-\theta_1) \, f_1$ and
$\langle 0 | J_{\mu5}^8|\eta_1\rangle
= \imath p_\mu \, \sin(\theta_8-\theta) \, f_8$.
The difference between $\theta_8$ and $\theta_1$ following from 
Eq.~(\ref{eqrel}) is analogous to the one derived within
ChPT\footnote{We
 like to emphasize that Eq.~(\ref{tas})
is {\em not} to be read as $|\eta\rangle = \cos\theta_8 |\eta_8\rangle
- \sin\theta_1 |\eta_1\rangle$ etc., i.e.\ Eq.~(\ref{eq3}) still holds.}
\cite{Leutwyler97}.

\section{Masses and Decay Constants}

The important relation that connects short-distance
properties, i.e.\ decay constants, with long-distance
phenomena, i.e.\ mass-mixing, is provided by the
divergences of axial-vector currents including the anomaly
($i=u,d,s,c,\ldots$)
\begin{equation}
\partial^\mu \, \bar q_i \, \gamma_\mu\gamma_5 \, q_i 
= 2 m_i \, j_5^i +
\frac{\alpha_s}{4\pi} \, G \tilde G , 
\label{anomaly}
\end{equation}
with $j_5^i = \bar q_i \, \imath\gamma_5 \, q_i$.
Taking matrix elements $\langle 0 | \ldots |P\rangle$ (for
instance $\langle 0 | \partial^\mu J_{\mu5}^s | \eta \rangle
= M_{\eta}^2 \, f_{\eta}^s$) and
using the definition of  the decay constants (\ref{eq5}),
the mass matrix in the quark flavor basis is fixed to have
the following structure \cite{FeKrSt98}
\beq
&&U(\phi) \, {\rm diag}[ M_\eta^2, M_{\eta'}^2] 
\, U^\dagger(\phi) \nonumber\\[0.1em]
&=&
\left( \begin{array}{cc}
m_{qq}^2 + 2 a^2 & \sqrt2 y a^2 \cr
\sqrt2 y a^2 & m_{ss}^2 + y^2 a^2
 \end{array} \right) 
\label{mass}
\eeq
with
\beq
m_{qq}^2 
&=& 2m_q \, 
\langle 0 |j_5^q | \eta_q\rangle/f_q 
\simeq M_\pi^2 , \nonumber \\[0.1em]
m_{ss}^2 &=& 2m_s \, 
\langle 0 |j_5^s | \eta_s\rangle/f_s 
\simeq 2 M_K^2 - M_\pi^2 
\eeq
and
\beq
&& a^2 = \frac{1}{\sqrt2 f_q} \, \langle 0 | \frac{\alpha_s}{4\pi}
\, G \tilde G| \eta_q \rangle , \qquad y= \frac{f_q}{f_s}
\eeq
The mass matrix in the octet-singlet basis can simply be
obtained from (\ref{mass}) by a rotation about the ideal
mixing angle.
Solving for $\phi,y,a^2$ and using $
f_q \simeq f_\pi$, $f_s \simeq \sqrt{2 f_K^2 - f_\pi^2}$,
one obtains the  ``theoretical'' values quoted in
Table~\ref{table}. 

Alternatively, the
mixing parameters can be determined from 
phenomenology without using the $SU(3)_F$ relations
for $m_{ii}^2$ and $f_i^2$.
The mixing angle $\phi$ can be determined
by considering appropriate ratios of decay widths/cross sections, in
which only the $\eta_q$ or $\eta_s$ component is probed, respectively.
The analysis 
of several independent
decay and scattering processes performed in \cite{FeKrSt98}
leads to $\phi = 39.3^\circ \pm 1.0^\circ$.
It is to be stressed that the so-obtained values for the mixing angle
$\phi$ (or equivalently for $\theta=\phi-\arctan\sqrt2$) are all
consistent with each other with a small experimental uncertainty
and agree with the  ``theoretical''  ones within 10\%.

With this value of the mixing angle the decay constants $f_q$ and
$f_s$ can be estimated from the $\eta,\eta'\to\gamma\gamma$ decay
widths\footnote{Note that again, the expressions for the
two-photon decay widths take the simple form only in the quark-flavor
basis, in which the decay constant matrix, 
appearing in the derivation of the anomalous decay, 
can be inverted in a trivial way.}
\begin{eqnarray}
\Gamma[\eta\phantom{'}\to\gamma\gamma] &=&
\frac{9\alpha^2 M_{\eta\phantom{{}'}}^3}{16 \pi^3}  
\left[\frac{C_{q} \cos\phi}{f_{q}} -
      \frac{C_{s}  \sin\phi}{f_{s}}\right]^2 \nonumber  \\[0.2em]
\Gamma[{\eta'}\to\gamma\gamma] &=&
\frac{9\alpha^2 M_{{\eta'}}^3}{16 \pi^3}  
\left[\frac{C_{q}  \sin\phi}{f_{q}} +
      \frac{C_{s}  \cos\phi}{f_{s}}\right]^2 \nonumber \\[0.1em] &&
\label{eq:gammapred}
\end{eqnarray}
where $C_q=5/9\sqrt2$ and $C_s=1/9$ are the proper charge factors.
Combined with the additional information from the structure
of the mass matrix, one obtains $f_q = (1.07 \pm 0.02) \, f_\pi$
and $f_s = (1.34 \pm 0.06) \, f_\pi$ (see also Table~\ref{table}).
Note that the corresponding
difference between $\theta_8,\theta,\theta_1$ (although
formally a higher order $SU(3)_F$ breaking effect) is enormous!

A prominent example which illustrates the difference between
the conventional approach
with $\theta_8=\theta=\theta_1$ and the present one
is given by the $J/\psi\to
P\gamma$ decays. Following \cite{Novikov:1980uy,BaFrTy95} the decay rates are 
proportional to the matrix elements $|\langle 0 |
\frac{\alpha_s}{4 \pi} \, G \tilde G | P \rangle|^2$ which
can be calculated using Eqs.~(\ref{anomaly},\ref{mass})
and $m_u\simeq m_d \simeq 0$, leading to
\beq
\frac{\Gamma[J/\psi\to\eta'\gamma]}{\Gamma[J/\psi\to\eta\gamma]}
&=& \tan^2\phi \, \frac{M_{\eta'}^4}{M_{\eta}^4} \,
\left(\frac{k_{\eta'}}{k_\eta}\right)^3 \cr &=&
\cot^2 \theta_8 \,  \left(\frac{k_{\eta'}}{k_\eta}\right)^3
\label{rjpsi}
\eeq
from which one obtains by comparison with
the experimental value \cite{PDG96}
 $\phi=39.0^\circ \pm 1.6^\circ$ (or
$\theta=-15.7^\circ \pm 1.6^\circ$) and 
$\theta_8 = -22.0^\circ \pm 1.2^\circ$.

Direct information on the decay constants $f_P^i$ can also be obtained
from the analysis of the form factors for $\gamma^*\gamma \to P$
at large photon virtualities, which are dominated by the valence
Fock states in (\ref{eq2},\ref{eq4}). Using the modified 
hard-scattering approach (see \cite{FeKr97b,FeKr98} and
references therein), again, 
the phenomenological parameter
set in Table~\ref{table} leads to a perfect description of the
experimental data \cite{CLEO97,L397}.

\begin{table*}[thb]
\caption{Theoretical and phenomenological values of
mixing parameters (for details, see \cite{FeKrSt98}).}
\label{table}
\begin{center}
\begin{tabular}{c| ccccc |ccccc }
& $f_{q}/f_\pi$ & $f_{s}/f_\pi$ & $\phi$ & $y$
& 
$a^2$ $[$GeV$^2]$
 & $f_8/f_{\pi}$ & $f_1/f_{\pi}$  & $\theta$ & $\theta_8$ & $\theta_1$   \\
\hline\hline
theory & 
 $\protect\phantom{\pm}1.00$ & $\protect\phantom{\pm}1.41$ & 
 $\protect\phantom{\pm}42.4^\circ$  & 
$\protect\phantom{\pm}0.78$ & $\protect\phantom{\pm}0.281$
 &$\protect\phantom{\pm} 1.28 $ & 
$\protect\phantom{\pm}1.15$ & $ -12.3^\circ$& $-21.0^\circ$ & $-2.7^\circ$ \\
phenom.   &
 $\protect\phantom{\pm}1.07$ & $\protect\phantom{\pm}1.34$ & 
 $\protect\phantom{\pm}39.3^\circ$ & 
 $\protect\phantom{\pm}0.81$ & $\protect\phantom{\pm}0.265$
  &$\protect\phantom{\pm} 1.26$ &
 $\protect\phantom{\pm}1.17$ & $-15.4^\circ$ & $-21.2^\circ$ & $- 9.2^\circ$ \\
%
\hline
\end{tabular}
\end{center}
\end{table*}

\section{$\eta-\eta'-\eta_c$ Mixing}

Since the derivation of the pseudoscalar mass matrix via
Eq.~(\ref{anomaly}) does not have to make use of flavor
symmetry, it can be generalized to $\eta-\eta'-\eta_c$
mixing in a straight forward manner \cite{FeKrSt98}, leading to 
a similar
mass matrix as in Eq.~(\ref{mass})
\beq
&& \left( \begin{array}{ccc}
m_{qq}^2 + 2 a^2 & \sqrt2 y a^2 & \sqrt2 z a^2 \cr
\sqrt2 y a^2 & m_{ss}^2 + y^2 a^2 & y z a^2 \cr
\sqrt2 z a^2 & y z a^2 & m_{cc}^2 + z^2 a^2
 \end{array} \right) 
\nonumber \\[0.1em] &&
\label{mass2}
\eeq
Of course the mixing between light and heavy pseudoscalars
is suppressed by the heavy masses, i.e.\
$a^2/m_{cc}^2$ may be treated as a small parameter, leading to
$m_{cc}^2 \simeq M_{\eta_c}^2$.
The second new parameter is also unambiguously fixed
$z=f_q/f_c \simeq f_q/f_{J/\psi}=0.35$.

From the phenomenological point of view, namely from the rather large
branching ratio for $B \to K \eta'$ reported by CLEO \cite{Behrens:1998dn},
one is mostly interested in the matrix elements of $\eta,\eta'$
with the charm axial-vector current
$
\langle 0 | \bar c \gamma_\mu \gamma_5 c |P\rangle = \imath
\, f_P^c \, p_\mu
$.
From the diagonalization of the mass matrix 
one obtains the following values 
\beq
&& f_{\eta\phantom{{}'}}^c = - f_c\,\theta_c\,\sin\theta_8 = (-2.4 \pm
0.2) \ {\rm MeV}, \cr &&
f_{\eta'}^c = \phantom{-} f_c\,\theta_c\,\cos\theta_8 = 
(-6.3 \pm 0.6) \ {\rm MeV} \cr &&
\eeq
where we have defined the mixing angle $\theta_c 
= - z \, \sqrt{2+y^2} \, a^2/M_{\eta_c}^2 \simeq -1.0^\circ$,
which is reasonably small
and in accord with Refs.~\cite{Ali97b,Petrov:1997yf,Chao:1989yp} and,
 in particular,
with the
independent bounds found from the analysis of the $\eta\gamma$ and
$\eta'\gamma$
transition form factors \cite{FeKr97b}. Obviously, the intrinsic
charm in $\eta'$ cannot induce a dominant contribution to
the $B\to K \eta'$ decays (via $b \to s c \bar c$), 
contrary to what is assumed occasionally
\cite{ChTs97,HaZh97}.

An immediate test of the parameter values is provided by
a similar ratio of $J/\Psi$ decay widths as in Eq.~(\ref{rjpsi}).
Most interestingly, via Eq.~(\ref{anomaly}), the intrinsic 
charm picture (i.e.\ $J/\psi \to c\bar c\gamma,\ c\bar c\to \eta'$)
and the gluon picture of ref.~\cite{Novikov:1980uy} turn out to
be equivalent
with the result \cite{FeKrSt98}
\begin{eqnarray}
\frac{\Gamma[J/\psi \to \eta'\gamma]}{\Gamma[J/\psi\to \eta_c\gamma]}
&=& \theta_c^2 \, \cos\theta_8^2 \,\,
\left(\frac{k_{\eta'}}{k_{\eta_c}}\right)^3 \cr
&=&  
\left(\frac{\langle 0 | \frac{\alpha_s}{4\pi} \, G \tilde G
    |\eta'\rangle}{\sqrt2 \, f_{\eta_{c}} \, M_{\eta_{c}}^2 } \right)^2 \,
\left(\frac{k_{\eta'}}{k_{\eta_c}}\right)^3 \cr &&
\label{etacetap}
\end{eqnarray}
The values of $\theta_c$ and $\theta_8$ found in our approach
perfectly reproduce the experimental value for this ratio \cite{PDG96}.

\section*{Acknowledgements}

I like to express my gratitude
to Peter Kroll and Berthold Stech for
a fruitful collaboration.
I further enjoyed valuable discussions with 
Hai-Yang Cheng, Alex Kagan, Alexey Petrov
and Vladimir Savinov.

\nocite{FeKr97b,FeKr98}

\end{document}